\documentclass[
    ,final            % use final for the camera ready runs
  ]
  {aipproc}

\layoutstyle{6x9}

\begin{document}

\title{Long lived central engines in Gamma Ray Bursts}

%<Replace this text with PACS numbers; choose from this list:
%                \texttt{http://www.aip..org/pacs/index.html}>}
%
% 95.30.Gv 	Radiation mechanisms; polarization
% 95.30.Jx 	Radiative transfer; scattering
% 95.85.Nv 	X-ray
% 95.85.Pw 	γ-ray
% 98.70.Rz 	γ-ray sources; γ-ray bursts
%

\author{Gabriele Ghisellini}{
  address={INAF -- Osservatorio Astronomico di Brera, Merate, Italy}
}

%\author{<author2>}{
%  address={<common address for author2 and author3>}
%}
%
%\author{<author3>}{
%  address={<common address for author2 and author3>}
%  ,altaddress={<author1 address>} % additional visiting address
%}
%

\begin{abstract}
The central engine of Gamma Ray Bursts may live much longer
than the duration of the prompt emission.
Some evidence of it comes from the presence of
strong precursors, post--cursors, and X--ray flares
in a sizable fraction of bursts.
Additional evidence comes from the fact that often the X--ray and 
the optical afterglow light curves do not track one another, 
suggesting that they are two different emission components.
The typical ``steep-flat-steep'' behavior of the X--ray light
curve can be explained if the same central engine responsible
for the main prompt emission continues to be active for a
long time, but with a decreasing power.
The early X--ray ``afterglow'' emission is then the
extension of the prompt emission, originating at approximately
the same location, and is not due to forward shocks.
If the bulk Lorentz factor $\Gamma$ is decreasing in time, the
break ending the shallow phase can be explained, since
at early times $\Gamma$ is large, and we see only a fraction
of the  emitting area. 
Later, when $\Gamma$ decreases, we see an increasing
fraction of the emitting surface up to the time when
$\Gamma \sim 1/\theta_{\rm j}$. 
This time ends the shallow phase of the X--ray light curve.
The origin of the late prompt emission can be the accretion
of the fall--back material, with an accretion rate 
$\dot M \propto t^{-5/3}$. 
The combination of this late prompt emission with the
flux produced by the standard forward shock can explain
the great diversity of the optical and the X--ray light curves.

\end{abstract}

\maketitle

%%%%%%%%%%%%%%%%%%%%%%%%%%%%%%%%%%%%%%%%%%%%
%% MAINMATTER
%%%%%%%%%%%%%%%%%%%%%%%%%%%%%%%%%%%%%%%%%%%%

\section{Introduction}

The X--ray light curves, as observed by 
{\it Swift}, have shown a complexity unforeseen before.
Besides the ``normal'' behavior, as observed by {\it Beppo}SAX
after several hours from the trigger, we now know that a good fraction
of GRBs show a steep decay soon after the end of the burst as
seen by BAT, followed by a plateau lasting for a few thousands
of seconds, ending at the time $T_A$ (following \cite{willi07}).
This behavior, named ``Steep--Flat--Steep" \cite{gt05, nousek05} 
% (e.g. Tagliaferri et al. 2005; Nousek et al. 2005)
has been interpreted in several ways (for reviews, see e.g. \cite{zhang07})
none of which seems conclusive.
Furthermore, in nearly half of the bursts, we observe X--ray flares,
of relatively short duration $\Delta t$ (i.e. $\Delta t/t\sim 0.1$,
see \cite{chinca}) occurring even several hours after the trigger.
Considering X--ray flares in different GRBs, \cite{lazzati08}
have shown that their average luminosity goes like $t^{-5/3}$,
the same time decay of the accretion rate of the fall--back material
(see \cite{chevalier89}; \cite{zhang08}).

The optical light curves are also complex, but rarely track the
behavior of the X--ray flux (see e.g. \cite{pana06, pana07}),
suggesting a possible different origin.
At the other time extreme, for 10--15\% of the bursts, we observe
precursor emission, separated from the main event, in some cases,
by hundreds of seconds.
The energy contained in these precursors is comparable to the
energy of the main event, and the spectrum is indistinguishable
\cite{burlon08}, suggesting that they are produced by the same engine
producing the main event.
%

% The spectral slope does not change across the temporal
% break from the shallow to the normal decay phase,
% ruling out a changing spectral break as a viable explanation.
% An hydrodynamical or geometrical nature of the break is instead preferred.

To explain the ``steep--flat--steep'' X--tray light curves, \cite{uhm07}
and \cite{genet07} (see also Daigne, these proceedings)
suggested that
the X--ray plateau emission is not due to the forward, 
but to the reverse shock running into ejecta of
relatively small (and decreasing) Lorentz factors.
This however requires an appropriate $\Gamma$--distribution of the
ejecta, and also the suppression of the X--ray flux produced by the 
forward shock. 

We \cite{gg07} instead suggested  
that the plateau phase of the X--ray (and sometimes of the optical) emission  
is due to a prolonged activity of the central engine (see also \cite{lp07}),
responsible for a ``late--prompt'' phase: 
after the early ``standard" prompt phase the central engine continues to 
produce for a long time (i.e. days) shells of progressively lower 
power and bulk Lorentz factor.
The dissipation process during this and the early phases 
occur at similar radii (namely close to the transparency radius).
The reason for the shallow decay phase, and for the break ending it,
is that the $\Gamma$--factors of the late shells are
monotonically decreasing, allowing to see an
increasing portion of the emitting surface, until all of it is visible.
Then the break occurs when $\Gamma(t)=1/\theta_j$.

% ------------------------------------------------------
\begin{figure}
\includegraphics[height=11cm, width=15cm]{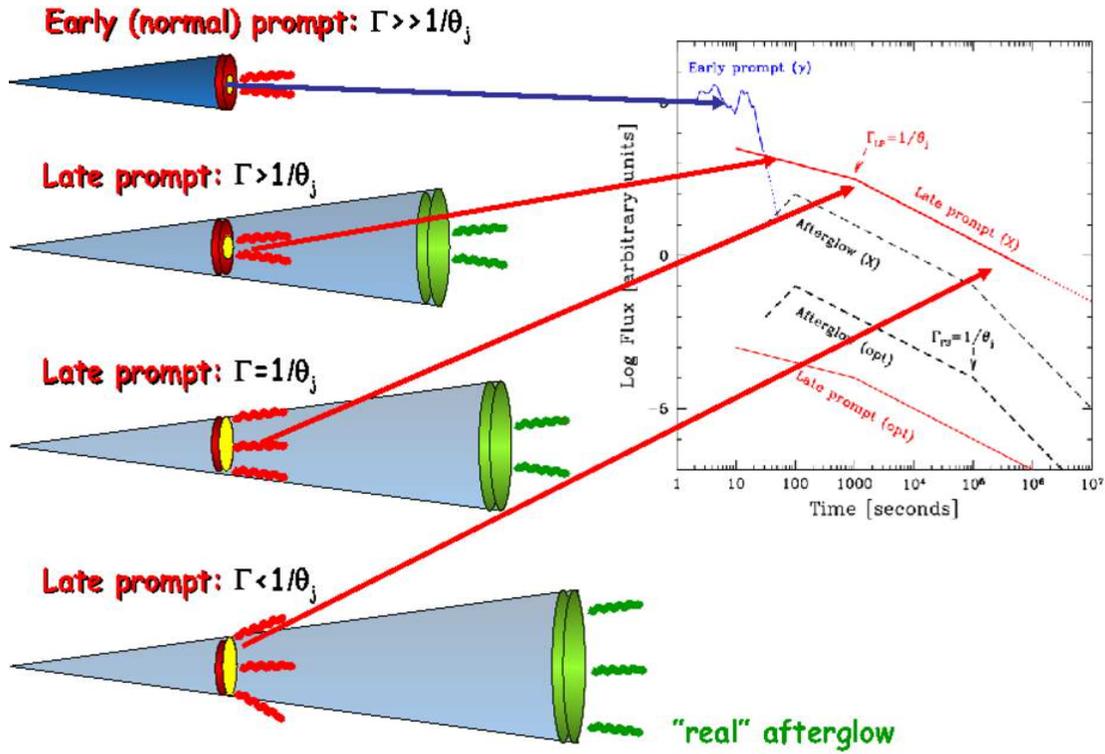}
\caption{
Cartoon of the proposed model, and schematic illustration of the different 
components contributing to the X--ray and optical light curves, as labeled.
Scales are arbitrary. 
The early prompt phase is erratic, with shells of varying $\Gamma$ and power. 
Then the central engine produces shells of progressively less power and
bulk Lorentz factors, producing a smoother light curve. 
Since the average $\Gamma$--factor is decreasing, the observer sees
an increasing portion of the emitting area, until all of it
becomes visible when $\Gamma \sim 1/\theta_j$.
When this occurs there is a break in the light curve,
associated with the ending of the shallow phase.
The case illustrated here is only one (likely the most common) 
possible case, when the X--ray flux is dominated by late 
prompt emission (solid line, the dotted line corresponds to an 
extrapolation at very late times), while the optical flux is dominated 
by the real afterglow (dashed). 
% $\Gamma_{\rm LP}$ and $\Gamma_{FS}$ indicate the $\Gamma$ of the late
% shells and the forward shocks, respectively. 
Adapted from \cite{gg07}.
}
\end{figure}
% ------------------------------------------------------

\section{The late prompt emission scenario}

Willingale et al. \cite{willi07} have proposed to describe the
X--ray afterglow light curve with a rising exponential connecting to
a power law function.
The end of the shallow phase is the junction between the exponential and
the power law, and it is called $T_A$.
% They showed that interpreting $T_A$ as a jet break time one
% obtains, for the {\it Swift} bursts in their sample, a good correlation
% between the peak energy of the prompt spectrum, $E_{\rm peak}$,
% and the collimation corrected energetics $E_\gamma$, with a small
% scatter and a slope identical to the so called Ghirlanda relation \cite{ggl04}
% (which identifies as a jet break time the break in the optical light curve,
% occurring usually much later), challenging the physical nature of
% the Ghirlanda relation.
% Nava et al. \cite{nava07} have then 
% investigated this issue with a larger sample, finding that the correlation
% found by \cite{willi07} does not have the same slope of the
% Ghirlanda one, and it is not as tight.
% More importantly, they demonstrated that $T_A$ does not play any role 
% in the construction of the correlation found by \cite{willi07},
% which is instead (entirely) a by--product of the the $E_{\rm peak}$--$E_{\rm iso}$
% correlation (the so called ``Amati" relation, \cite{ama02}).
% In fact there is no (anti)--correlation between $T_a$ and $E_{\rm iso}$
% (``a la Frail", \cite{frail01}) for GRBs of the same $E_{\rm peak}$
% (see \cite{nava07} for more details and figures).

Investigating the possibility that $T_A$ might be a jet break, as suggested by
\cite{willi07}, we \cite{nava07} demonstrated that it is not, yet 
it may be produced by a 
mechanism similar to the process responsible for the
jet break visible during the deceleration of the fireball.
Suppose that the accretion onto the newly formed 
black hole occurs in two phases: 
the first is short, intense, erratic,
corresponding to the early prompt phase of GRBs.
The accreting matter may be the equatorial core material 
which failed to form the black hole in the first place. 
Being very dense, it can sustain a strong magnetic
field, which in turn efficiently extracts the rotational energy of the
black hole.  
The second phase is longer, smoother, with a rate decreasing in time,
corresponding to the late prompt emission.
The accreting matter may be fall--back material,
with a density  smaller than in the early phases. 
The magnetic field that this matter can sustain is weaker than
before, with a corresponding smaller power extracted from the black hole spin.
This may well correspond to production of shells of smaller 
$\Gamma$--factors.
These shells can dissipate part of their energy with the same mechanism
of the early ones. 
Occasionally, in this late prompt phase,
the central engine may produce a faster than average shell,  
originating the late flares often observed in the Swift/XRT light curves.

In this scenario there is a simple relation between time profile
of $\Gamma$ and the observed decay slopes before and after $T_A$.
The plateau phase is described by $L(t)\propto t^{-\alpha_2}$,
followed by a steeper decay $L(t)\propto t^{-\alpha_3}$.
Then, by geometry alone, \cite{gg07} derived:
\begin{equation}
\Gamma \, \propto  t^{-(\alpha_3-\alpha_2)/2}
\end{equation}
By setting $L(t) = \eta \Gamma \dot M_{\rm out} c^2 \propto t^{-\alpha_3}$
(after $T_A$ we see the entire jet surface) we get:
\begin{equation}
\dot M_{\rm out} \,  \propto \, t^{-(\alpha_2+\alpha_3)/2}
\end{equation}
Since $\alpha_3 >1$, the total energy involved in the late prompt
emission {\it is modest, at most comparable with the entire early prompt energy}.

% ------------------------------------------------------
\begin{figure}
\includegraphics[height=14cm, width=15cm]{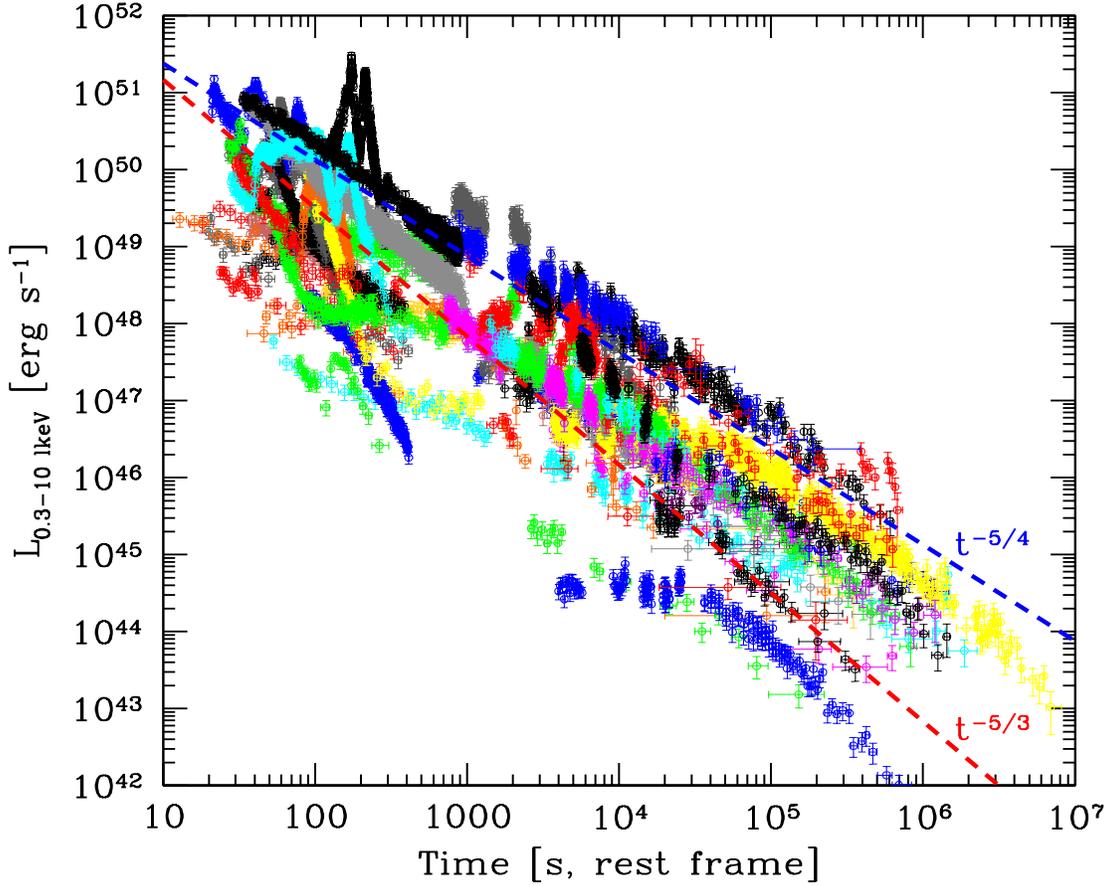}
\caption{
The X--ray light curve (in the 0.3--10 keV band)
for a sample of 33 {\it Swift} GRBs with redshift and measured
extinction on the optical. The two dashed line correspond
to $t^{-5/4}$ and $t^{-5/3}$: the latter appears to be too steep
to explain the observed general behavior. 
But when the light curves are de--convolved into the late prompt
and ``real'' afterglow components, we recover $t^{-5/3}$ for the
late prompt emission after the end of the shallow phase.
}
\label{total}
\end{figure}
% ------------------------------------------------------

%
%
%% ------------------------------------------------------
\begin{figure}
\begin{tabular}{cc}
\includegraphics[height=9.cm, width=7.5cm]{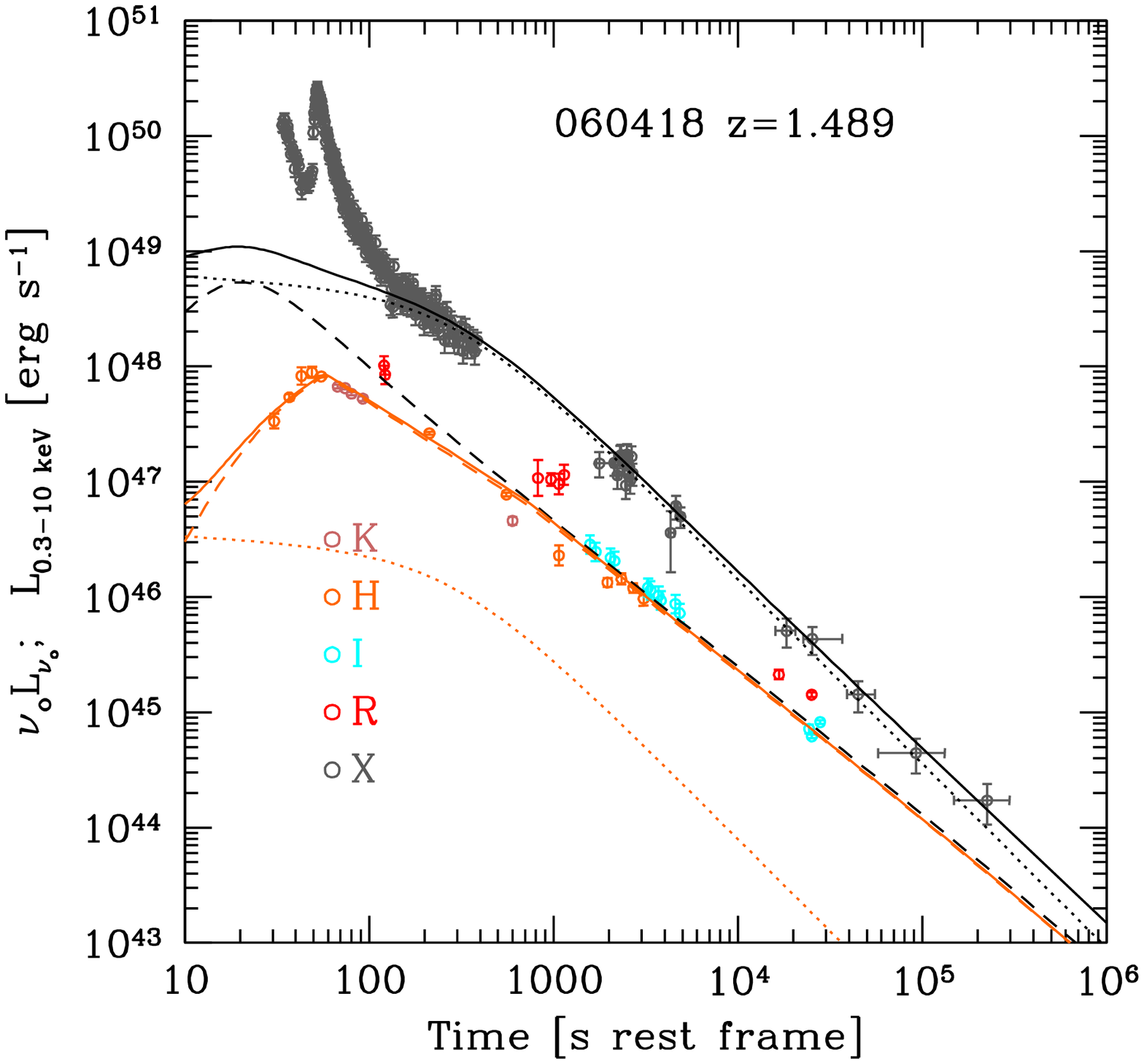}
& \includegraphics[height=9.cm, width=7.5cm]{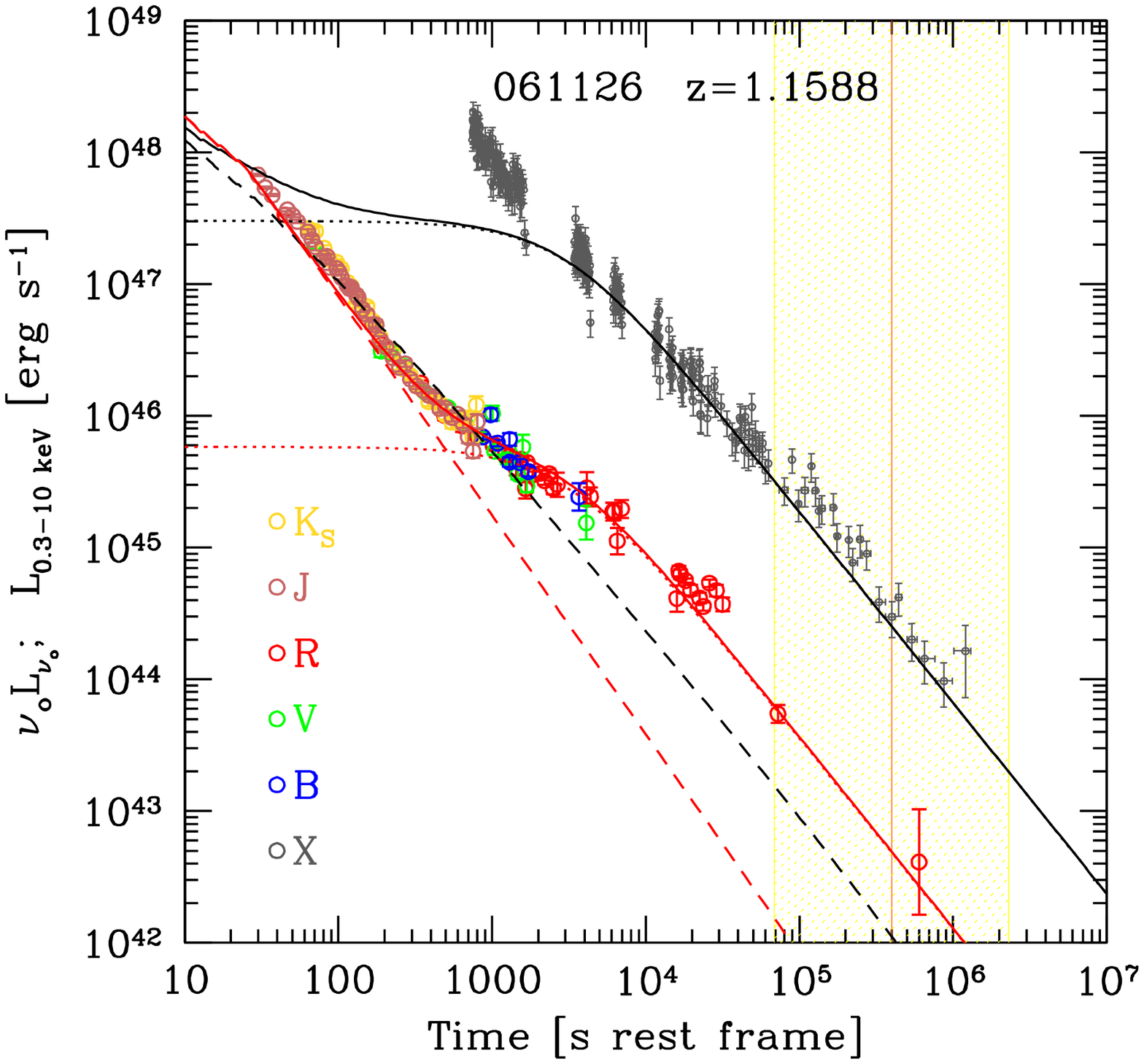}\\
\includegraphics[height=9.cm, width=7.5cm]{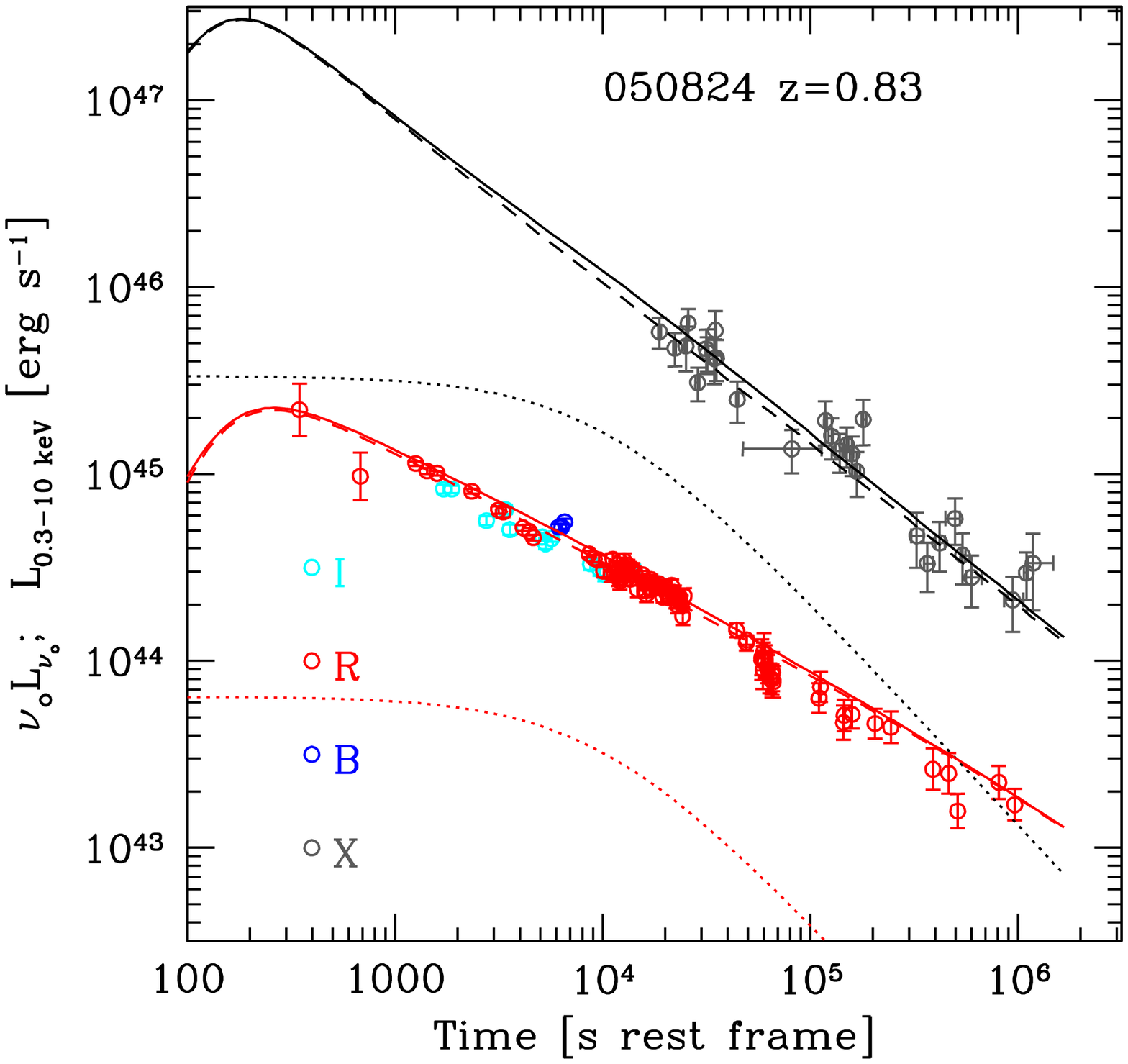}
&\includegraphics[height=9.cm, width=7.5cm]{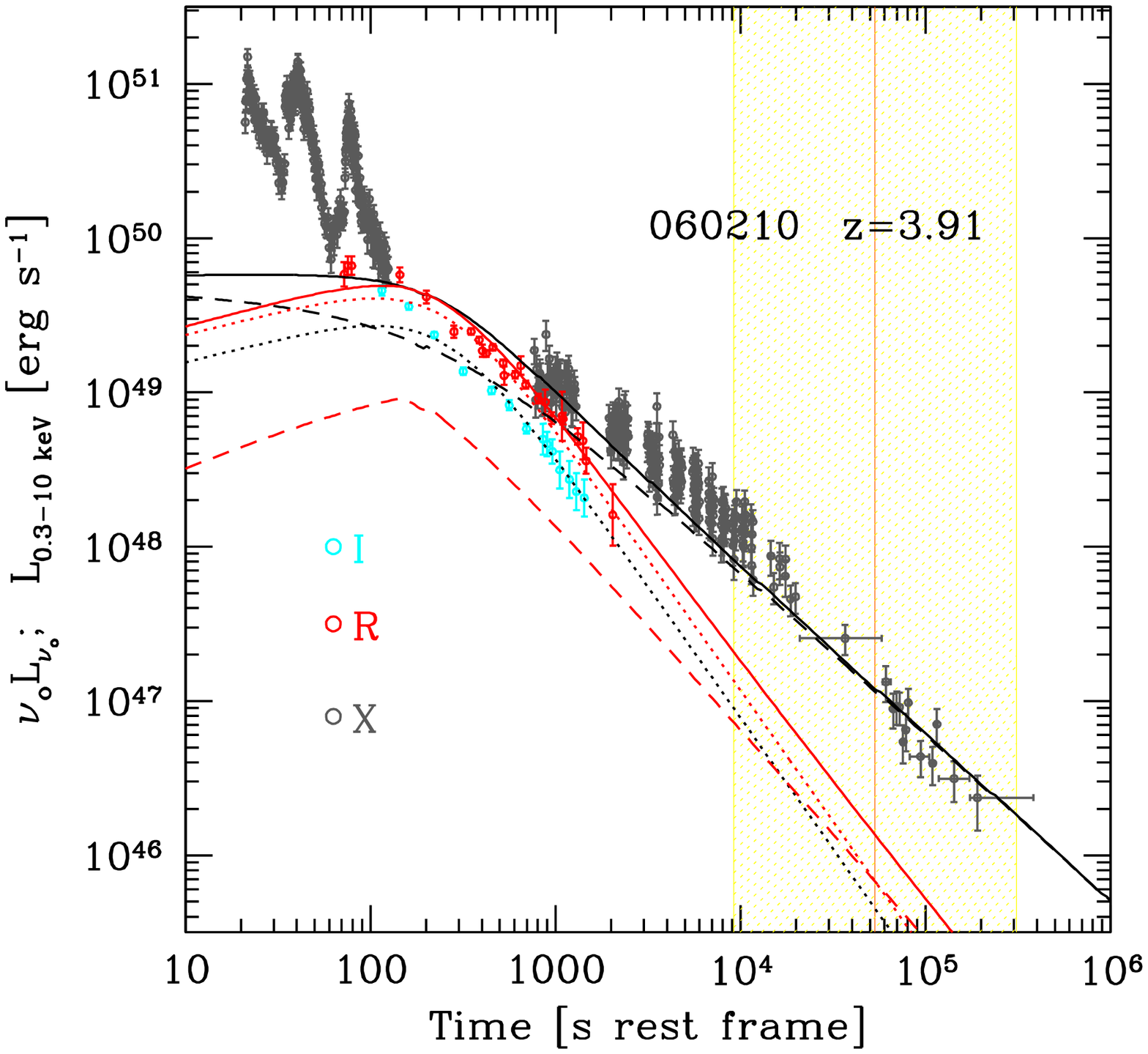}
\end{tabular}
\caption{Some examples of X--ray [0.3--10 keV] and optical 
($\nu L_\nu$) light--curves, calculated in the rest frame.
The optical and X--ray luminosities  have been de--absorbed and K--corrected.
The vertical line (and shaded band)
is the rest frame jet break time (and its 3$\sigma$ uncertainty) expected 
if the burst obeys the Ghirlanda relation \cite{ggl04}.
The dashed and dotted lines correspond to ``real" afterglow and late prompt
emission, respectively. Solid lines correspond to the sum.
In GRB 060418 (representative of the most common case), the late prompt process
dominates the X--ray luminosity, and the real afterglow dominates the optical one.
In GRB 061126, instead, the optical at early times is dominated
by the real afterglow, but becomes late prompt emission at later times.
For GRB 050824 the real afterglow emission dominates in both bands.
Finally, in GRB 060210 we have that the real afterglow dominates the X--ray
luminosity, while the late prompt is important in the optical band, at least
for the plotted (early) times (from \cite{gg09}).
}
\label{combo4}
\end{figure}
% ------------------------------------------------------

% If we insert the average values of $\alpha_3$ and $\alpha_2$ 
% ($\sim 1.25\pm 0.25$ and $\sim 0.6\pm 0.3$, respectively, see \cite{pana06})
% we approximately have $\dot M\propto t^{-1}$ and $\Gamma\propto t^{-1/3}$.
% This means that the total energy (i.e. integrated over time, 
% $E =\int \Gamma \dot M c^2 dt$, beginning from the start of the plateau phase)
% involved in the late phase is smaller than the energy spent during
% the early prompt.

\section{A unifying view} 

The late prompt emission scenario {\it adds} one component, 
it does not substitute the forward shock emission with something else.
Both mechanisms are working, 
both making some X--ray and optical flux.
We may then have a variety of cases: both the optical and the X--rays
can be late prompt emission or forward shock emission;
or X--rays and optical can be ``decoupled'',  
one due to late prompt and the other to the forward shock (that hereafter
we call ``real afterglow'').

We \cite{gg09} have then tested these ideas collecting all X--ray and optical light--curves
of {\it Swift} bursts of known redshift and optical extinction (at the host).
This ensures that we can construct reliable luminosity vs rest frame time
profiles for both bands.
Fig. \ref{total} shows all the X--ray light--curves in the rest frame.
Overall, they appear to decay roughly as $t^{-5/4}$ (see the appropriate
dashed line).
This is slope is flatter than $t^{-5/3}$, expected if the 
the X--ray luminosity is proportional to the the accretion 
rate of the fall--back material (of the massive exploding star) 
onto the black hole (\cite{chevalier89}, \cite{zhang08}).

For all the 33 GRBs we tried to model both 
the X--ray and the optical light curve with the sum of two components.
The first is the standard forward shock emission, calculated following the
prescription of \cite{pana00}, the second mimics the late prompt emission 
through a phenomenological parametrization.
We have then doubled the number of input parameters (6 for the forward shock,
7 for the ``late prompt"), and the fact that we obtain
a good representation of the data may come not as a surprise.
However, several parameters of the phenomenological (late prompt) part
are really well constrained, such as the time $T_A$, the time decay slopes before and
after $T_A$, and the X--ray frequency spectrum. 
We obtain {\it in all cases} a good representation.
Fig. \ref{combo4} shows 4 examples, selected to show 4 different cases.
GRB 060418 illustrates the most common case: the late prompt process
dominates the X--ray luminosity, and the ``real afterglow'' dominates the optical one.
In GRB 061126, instead, the optical at early times is dominated
by the real afterglow, but becomes late prompt emission at later times.
For GRB 050824 the real afterglow emission dominates in both bands.
Finally, in GRB 060210 we have that the real afterglow dominates the X--ray
luminosity, while the late prompt is important in the early optical band
(where we have data).
The main outcomes of this study are: i) we can explain all the X--ray and optical
light--curves within a relatively simple scenario; ii) we can understand
why it is difficult to have achromatic jet breaks (but sometimes we do);
iii) even when we can see the jet break, the flux after the jet break time
can decay in a shallower way than predicted
(the unbroken late prompt emission can still contribute);  iv) the 
decay slope of the X--ray late prompt emission can indeed be $5/3$ 
even if we observe a flatter slope, since at relatively late times
the real afterglow emission helps in flattening the light curve.

\begin{theacknowledgments}
I gratefully thank all my collaborators: D. Burlon, A. Celotti, C. Firmani, G. Ghirlanda,
D. Lazzati, M. Nardini, L. Nava and F. Tavecchio.
\end{theacknowledgments}



% \bibliographystyle{aipproc}   % if natbib is available
\bibliographystyle{aipprocl} % if natbib is missing

% \bibliography{sample}

% \IfFileExists{\jobname.bbl}{}
%  {\typeout{}
%  \typeout{******************************************}
%  \typeout{** Please run "bibtex \jobname" to optain}
%  \typeout{** the bibliography and then re-run LaTeX}
%  \typeout{** twice to fix the references!}
%  \typeout{******************************************}

%%%%%%%%%%%%%%%%%%%%%%%%%%%%%%%%%%%%%%%%%%%
%% The following lines show an example how to produce a bibliography
%% without the help of the BibTeX program. This could be used instead
%% of the above.
%%%%%%%%%%%%%%%%%%%%%%%%%%%%%%%%%%%%%%%%%%%

% \endinput
\end{document}